\documentstyle[aps,epsf,floats,twocolumn,prl]{revtex}
\begin{document}

\twocolumn[\hsize\textwidth\columnwidth\hsize\csname@twocolumnfalse%
           \endcsname

\title{Localization on short-range potentials \\
       in dissipative quantum mechanics}

\author{Akakii Melikidze}
\address{Physics Department, Princeton University, Princeton, NJ
         08544}

\date{\today}

\maketitle

\begin{abstract}
In this article the problem of the existence of a state localized on a
weak short-range attractive potential in the presence of dissipation is
considered. It is shown that, contrary to the pure quantum case, a
localized state is produced in any number of dimensions, while in low
dimensions dissipation leads to much stronger localization. The results
have physical implications for the dissipative dynamics of objects such as
heavy particles in Fermi liquids and for superconductivity in high-$T_c$
materials.

\end{abstract}


]


Research in the theory of strongly interacting fermionic systems was
rekindled soon after the discovery of high temperature cuprate
superconductors when it was realized that Landau Fermi liquid theory can
not describe the normal state of the underdoped materials. Since then,
many ideas have been proposed to exhibit possible routes to non-Fermi
liquid behavior.
\cite{Anderson_book,Varma,Lee,SF_theories,Sachdev_Science,Schofield}
While these theories
differ in details, they all are intimately related to a peculiar property
of fermionic liquids known as Anderson's orthogonality
catastrophe.~\cite{Anderson_catastrophe} This property implies that the
wave function of the liquid before and after an adiabatic insertion of an
impurity has a vanishing overlap:
\begin{equation}
\label{Overlap_Anderson}
\langle \psi | \psi'\rangle \sim N^{-A},
\end{equation}
where $N$ is the number of fermions in the liquid and $A$ is the
orthogonality exponent which can be expressed in terms of impurity
scattering phase shifts. This result shows that interactions between
fermions have singular effect and the fact that the non-interacting
quasiparticles in Landau theory can often be defined is rather an
exception than a principle.

Kondo~\cite{Kondo} introduced a closely related overlap of the wave
function of the liquid with an impurity at 0 and that with the
impurity moved slowly to ${\bf x}$:
\begin{equation}
\label{Overlap_Kondo}
\langle \psi(0) | \psi({\bf x})\rangle
   \sim (\Lambda t)^{-K({\bf x})},
\end{equation}
where $t$ is a characteristic time of the impurity motion and $\Lambda$ is
a high energy cut-off. The exponent $K({\bf x})$ is quite difficult to
evaluate in general; however, it was shown by 
Sch\"onhammer~\cite{Schonhammer} that for small ${\bf x}$ ($\hbar=1$
throughout the letter):
\begin{equation}
\label{Exponent_Schonhammer}
K({\bf x}) \approx \frac{\eta {\bf x}^2}{2\pi}.
\end{equation}
Here $\eta$ is the friction coefficient that determines the viscous force
${\bf F}=-\eta \dot{{\bf x}}$ exerted on the moving impurity by the
fermions in the liquid. That there is such a force can be easily seen by
making a Galilean transformation from the reference frame of the liquid to
that of the impurity. There, the fermionic liquid flows past the impurity
and the net force produced by scattering on the impurity has the same
origin as the well-known residual resistance of metals. Moreover, this
transformation allows a simple calculation of $\eta$ which can be
expressed in terms of the impurity's transport scattering
cross-section.~\cite{Weiss}

The significance of Eq.~(\ref{Exponent_Schonhammer}) is that it relates an
adiabatic effect --- orthogonality catastrophe --- with an essentially
non-adiabatic effect --- energy dissipation caused by the viscous
force. In
problems where the particle of interest can be singled out (like, for
example, diffusion of heavy particles in Fermi liquids, the dynamics of
topological defects, solitons, collective coordinates etc.) this
connection provides a powerful tool that reduces the original
strongly interacting problem to what is known as dissipative quantum
mechanics of a single particle.~\cite{Weiss} As for the Fermi liquid
itself, one can not, in general, deduce the breakdown of Landau theory
using these simple arguments. However, once the breakdown occurs, its
direct cause --- Anderson's orthogonality catastrophe --- can be linked to
the
dissipative dynamics of fermions. In the case of cuprate superconductors
this idea is supported by experimental evidence of strongly damped
quasiparticles.~\cite{Orenstein_Millis}


Motivated by these connections, I study the problem of localization of a
particle on attractive short-range potentials in $D$ dimensions in the
presence of Ohmic dissipation at zero temperature. This problem is
relevant to the localization of heavy particles on impurities in fermion
liquids, pinning of vortices in superconductors~\cite{Melikidze} etc. It
is also relevant to the problem of interaction between dissipative
particles and to superconductivity. The question that I address is
whether a short range attractive potential leads to localization of the
particle and, if so, what are the properties of the localized state. I
find that localization occurs for arbitrarily weak attractive potentials
in any number of dimensions and for any strength of dissipation. This
should be compared to the corresponding problem in the non-dissipative
case~\cite{Landau_Lifshits} where it is known that weak potentials
localize the particle in $D=1$, lead to exponentially weak localization in
$D=2$ and do not lead to localization in $D\ge 3$. I suggest that the much
stronger localization in the dissipative case may be relevant to the
physics of Anderson localization and the problem of high critical
temperature in cuprate superconductors.

Consider a particle of mass $m$ that moves in $D$ dimensions in the
presence of an attractive short-range potential
$V({\bf x})=-\alpha\delta({\bf x})$. The particle also experiences
the action of a friction force that is linear in the particle's velocity:
${\bf F}=-\eta \dot{{\bf x}}$ (Ohmic case). The well-accepted
microscopic approach to this problem is to think of the friction force as
originating from coupling to an effective environment. The environmental
degrees of freedom are eventually traced over to produce an effective
description of the particle's dynamics. I use the imaginary (single) time
path integral representation of the partition function.~\cite{Feynman}
This way, tracing over the environmental degrees of freedom reduces to
path-integration over environmental variables, yielding the following
effective action description:~\cite{Caldeira_Leggett}
\begin{eqnarray}
\label{Z}
Z &=& \int {\cal D}{\bf x}\, e^{-S},\\
\label{S}
S &=& \frac{T}{2} \sum_{n=-\infty}^{+\infty}
  \left( m\nu_n^2 + \eta|\nu_n| \right) |{\bf x}_n|^2\nonumber\\
  &-&\alpha \int_{0}^{1/T} d\tau\, \delta({\bf x}(\tau)).
\end{eqnarray}
Here $\nu_n=2\pi Tn$ are Matsubara frequencies:
${\bf x}(\tau)=\sum e^{i\nu_n\tau}{\bf x}_n$.

The $\eta|\nu_n|$ term in the effective action $S$ is responsible for
dissipation. Note that it has a non-analytic $|\nu_n|$ frequency
dependence. This fact strongly complicates the problem because it leads to
a non-local effective action when written in time representation.
However, there are two limits in which the problem is somewhat
simplified: the pure quantum limit and the dissipative limit. Indeed, the
interplay between different terms in $S$ produces a characteristic energy
scale $\omega_0$. Then, in the pure quantum limit $\omega_0\gg\eta/m$ the
dissipative term in $S$ is much smaller than the kinetic energy
term and can, therefore, be neglected. The resulting pure
quantum problem can be solved exactly.~\cite{Landau_Lifshits} In the
opposite dissipative limit $\omega_0\ll\eta/m$ the kinetic energy term can
be neglected; its only purpose then is to set a high-energy cut-off.

To analyze the problem I employ the variational method.\cite{Feynman} This
method exploits the fact that for an arbitrary trial action $S_0$ the
following inequality puts an upper bound on the free energy:
\begin{eqnarray}
\label{Theorem}
F \le F_0 + T\langle S-S_0\rangle_{S_0}.
\end{eqnarray}
Here $F = T\ln Z = T\ln\int{\cal D}{\bf x}\,e^{-S}$, 
$F_0 = T\ln Z_0 = T\ln\int{\cal D}{\bf x}\,e^{-S_0}$ and
$\langle S-S_0\rangle_{S_0} = Z_0^{-1}\int{\cal D}{\bf
x}\,(S-S_0)e^{-S_0}$.
We are interested in the ground state energy: $E=\lim_{T\to 0}F$.
Taking the low temperature limit in Eq.~(\ref{Theorem}) we get:
\begin{eqnarray}
\label{Low_T}
E \le E_0 + \lim_{T\to 0} T\langle S-S_0\rangle_{S_0}.
\end{eqnarray}
Here $E_0=\lim_{T\to 0}F_0$. I take for $S_0$ the action of an
Ohmic-damped harmonic oscillator:
\begin{eqnarray}
\label{S_0}
S_0 = \frac{T}{2} \sum_{n=-\infty}^{+\infty}
      \left( m\nu_n^2 + \eta|\nu_n| + m\omega_0^2 \right) |{\bf x}_n|^2.
\end{eqnarray}
Here, the $\omega_0$ of the oscillator will serve as a variational
parameter. As we shall see below, all quantities that are necessary for
the evaluation of Eq.~(\ref{Low_T}) can be expressed through the
expectation values of the operators that are diagonal in the
$x$-representation. This means that we only need to know the diagonal
elements of the equilibrium density matrix $\rho_0$ that corresponds to
$S_0$. However, the fact $S_0$ is quadratic implies that $\rho_0$ is
Gaussian. Thus, in the $T\to 0$ limit we can write:
\begin{eqnarray}
\label{Rho_0}
\rho_0({\bf x},{\bf x})
  =\frac{1}{(2\pi\langle x^2\rangle_0)^\frac{D}{2}}
   \exp\left\{-\frac{{\bf x}^2}{2\langle x^2\rangle_0}\right\}.
\end{eqnarray}
Here, $\langle x^2\rangle_0=\frac{1}{D}\langle {\bf x}^2\rangle_0$
is the ground state expectation value of $x^2$ for the damped harmonic
oscillator that corresponds to $S_0$. This quantity can be easily
evaluated:
\begin{eqnarray}
\label{X_def}
\langle x^2\rangle_0
 = \lim_{T\to 0} T\sum_{n=-\infty}^{+\infty}
      \frac{1}{m\nu_n^2+\eta|\nu_n|+m\omega_0^2}.
\end{eqnarray}
In the $T\to 0$ limit the sum can be substituted by an integral. Defining
$r=2m\omega_0/\eta$, we can write the resulting expression as:
\begin{eqnarray}
\label{X_result}
\langle x^2\rangle_0
= \frac{1}{\pi\eta}\times
\cases{
  \frac{1}{\sqrt{1-r^2}}\ln\frac{1+\sqrt{1-r^2}}{1-\sqrt{1-r^2}},
  &$r<1$;\cr
  \frac{2}{\sqrt{r^2-1}}\arctan\sqrt{r^2-1},
  &$r>1$.\cr
} 
\end{eqnarray}
In particular, in the dissipative regime $r\ll 1$ we get:
$\langle x^2\rangle_0=(2/\pi\eta)\ln(1/r)$, while in
the pure quantum regime $r\gg 1$ we obtain the correct quantum
expression: $\langle x^2\rangle_0=1/2m\omega_0$; at $r=1$ $\langle
x^2\rangle_0$ is analytic and equals $2/\pi\eta=1/\pi m\omega_0$. Note
that $\langle x^2\rangle_0$ for $r>1$ can be obtained from the expression
for $r<1$ by an analytic continuation.

Next, using Eq.~(\ref{Rho_0}) we can evaluate:
\begin{eqnarray}
\label{S-S_0}
\lim_{T\to 0} T\langle S-S_0\rangle_0
  &=& \langle-\alpha\delta({\bf x})
      -\frac{m\omega_0^2}{2}{\bf x}^2\rangle_0\nonumber\\
  &=& -\frac{\alpha}
       {\left(2\pi\langle x^2\rangle_0\right)^\frac{D}{2}}
      - \frac{Dm\omega_0^2}{2}\langle x^2\rangle_0.
\end{eqnarray}
To evaluate $E_0$ we apply Hellmann-Feynman theorem:
$\partial E_0/\partial\omega_0 = m\omega_0\langle {\bf x}^2\rangle_0$;
this leads to:
\begin{eqnarray}
\label{E_0}
E_0 = Dm\int_0^{\omega_0}d\omega_0'\,\omega_0'\langle x^2\rangle_0.
\end{eqnarray}
Substituting Eqs.~(\ref{S-S_0},\ref{E_0}) into Eq.~(\ref{Low_T}) and
integrating by parts, we get a variational bound on the ground state
energy:
\begin{eqnarray}
\label{E_w}
E \le
- \frac{Dm}{2}\int_0^{\omega_0}d\omega_0'\,\omega_0'^2
  \frac{\partial}{\partial\omega_0'}\langle x^2\rangle_0
  - \frac{\alpha}{\left(2\pi\langle x^2\rangle_0\right)^\frac{D}{2}}.
\end{eqnarray}

An immediate consequence of this result is the following statement: For
any combination of non-zero values of dimensionality $D$, mass of the
particle $m$ and friction constant $\eta$ the ground state energy
$E<0$. And since for any delocalized state $E\ge 0$ this implies that the
particle is always localized! To see this, notice that for small enough
values of $\omega_0$ it follows from Eq.~(\ref{X_result}) that the first
term in Eq.~(\ref{E_w}) is positive and $\propto \omega_0^2$
while the second term is negative and its absolute value is $\propto
(\ln\frac{1}{\omega_0})^{-D/2}$. Therefore for sufficiently small
$\omega_0$ the second term dominates and the energy of the ground state is
bounded from above by a negative value.

Minimizing Eq.~(\ref{E_w}) with respect to $\omega_0$ we obtain an
equation on $\omega_0$:
\begin{eqnarray}
\label{Omega_0}
\frac{\alpha}{(2\pi)^\frac{D}{2}} = m\omega_0^2\langle
x^2
\rangle_0^{\frac{D}{2}+1}.
\end{eqnarray}
The upper bound on the ground state energy can be found numerically by
solving Eqs.~(\ref{E_w},\ref{Omega_0}); this solution is presented
below. However, in two limiting cases, $\omega_0\ll\eta/m$ and
$\omega_0\gg\eta/m$, approximate analytical expressions can be
obtained. These two limits are realized for weak and strong pinning
potentials respectively. To be more precise, let us define $\alpha_c$ as
the strength of the potential at which the minimizing value $\omega_0$
satisfies $r=2m\omega_0/\eta=1$. Substituting this in
Eqs.~(\ref{X_result},\ref{Omega_0}) we obtain:
\begin{eqnarray}
\label{Alpha_c}
\alpha_c = \frac{2^{D-1}}{\pi m}\eta^{1-\frac{D}{2}}.
\end{eqnarray}

For weak attractive potentials, such that $\alpha\ll\alpha_c$, we have
the dissipative regime: $\omega_0\ll\eta/m$. In this case, using the
limiting form of Eq.~(\ref{X_result}) for $r\ll 1$ and solving
Eqs.~(\ref{E_w},\ref{Omega_0}) we obtain:
\begin{eqnarray}
\label{E_diss}
E_{\rm diss} \le -\frac{\eta}{2\pi m}\left(\frac{\alpha}{\alpha_c}\right)
\left(\ln\frac{\alpha_c}{\alpha}\right)^{-\frac{D}{2}}.
\end{eqnarray}
This result shows that the localization on sufficiently weak attractive
potentials is dominated by dissipative dynamics and is much stronger
compared to the non-dissipative case. Indeed, note the linear
dependence of $E_{\rm diss}$ on $\alpha$ in Eq.~(\ref{E_diss}) vs.
exact results in the pure quantum case:\cite{Landau_Lifshits}
$E_{\rm quant}= -m\alpha^2/2$ in $D=1$;
$E_{\rm quant}=-(m\lambda^2)^{-1}\exp(-2\pi/m\alpha)$ in $D=2$
($\lambda$ is the short-distance cut-off); no localization on weak
potentials in $D\ge 3$.

In the opposite limit of sufficiently strong potentials, such that
$\alpha\gg\alpha_c$, we have the pure quantum regime:
$\omega_0\gg\eta/m$. Here the situation depends on the dimensionality. In
$D=1$, using the limiting form of Eq.~(\ref{X_result}) for $r\gg 1$ and
solving Eqs.~(\ref{E_w},\ref{Omega_0}), we obtain:
\begin{eqnarray}
\label{E_quant}
E_{\rm quant}^{(D=1)} \le - \frac{m\alpha^2}{\pi}.
\end{eqnarray}
This bound is reasonably close to the exact quantum-mechanical result
$E=-m\alpha^2/2$. In dimensions $D\ge 2$ the pure quantum problem of a
strong ($\alpha\gg\alpha_c$) attractive potential acquires a short
distance cut-off (bandwidth) dependence; this is a direct consequence of
the fact that $\delta$-function is poorly defined in $D>1$. However, it
should be emphasized that the particle is always localized in any $D$ (see
the arguments above)! While the question about the ground state energy in
the strong potential case lies outside the main line of this letter, it
could be mentioned that, in the first approximation, 
$E(\alpha\gg\alpha_c)$ is bounded from above by
$E_{\rm diss}(\alpha\sim\alpha_c)\sim -\eta/m$ for any $D$.

\begin{figure}
\epsfxsize=\columnwidth
\centerline{\epsffile{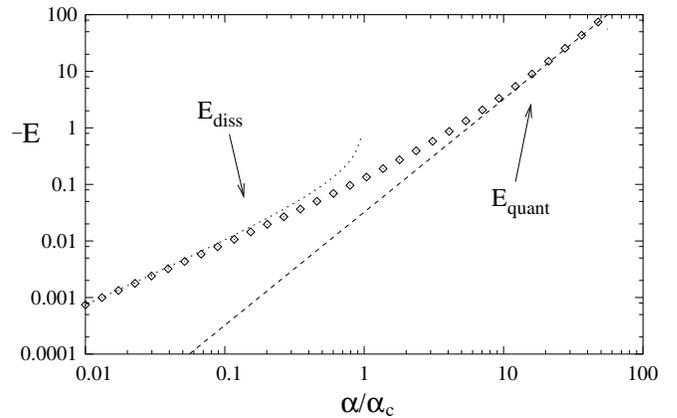}}
\medskip
\caption{Numerical minimization of Eq.~(\ref{E_w}) (diamonds) shows the
dependence of the upper bound on the ground state energy $E$ (in units of
$\eta/m$) on the strength of the potential $\alpha/\alpha_c$ for $D=1$.
Also shown are $E_{\rm diss}$, Eq.~(\ref{E_diss}) (dotted line) and
$E_{\rm quant}$, Eq.~(\ref{E_quant}) (dashed line).}
\label{Fig1}
\end{figure}

Finally, let us discuss the numerical minimization of Eq.~(\ref{E_w}). The
results of the solution for $D=1$ are presented in Fig.~\ref{Fig1}. A
crossover between the $\propto \alpha^2$ behavior in the pure quantum
regime (large $\alpha$) and the $\propto \alpha$ behavior in the
dissipative regime (small $\alpha$) occurs at $\alpha\sim\alpha_c$. Note
the much stronger localization that occurs for weak potentials compared to
the pure quantum case. One can see that the two limiting expressions,
Eqs.~(\ref{E_diss},\ref{E_quant}) describe the numerical results well in
their respective limits.

To understand the qualitatively different localization in the dissipative
case let us first recall how localization occurs in the non-dissipative
case. There, it is determined by the competition of two tendencies of the
particle: to localize and thus lower its potential energy and to diffuse
away thus lowering its kinetic energy. In the dissipative case, however,
the diffusion is strongly suppressed by the Anderson's orthogonality
catastrophe (see Eq.~(\ref{Overlap_Kondo})). Indeed, as was shown by Hakim
and Ambegaokar \cite{Hakim_Ambegaokar}, the mean-squared
displacement of a free Ohmic-damped particle at $T=0$ (this is known as
quantum Brownian motion problem) grows very slowly with time:
\begin{equation}
\label{Mean_squared}
\langle x^2\rangle = \frac{2}{\pi\eta}\ln\frac{\eta t}{m}.
\end{equation}
(This can also be shown~\cite{Weiss} to be a direct consequence of the
fluctuation-dissipation theorem of Callen and 
Welton~\cite{Callen_Welton}). Thus, strictly speaking, the diffusion
coefficient of the Ohmic-damped particle is zero. Therefore, the tendency
to lower potential energy by localization dominates over diffusion and
leads to localization for any $D$, in contrast to the pure
quantum case. Meanwhile, in one dimension dissipation leads to much
stronger localization on weak potentials compared to the non-dissipative
case.

Physical implications of these results are quite interesting. For example,
if one considers the problem of diffusion of a heavy particle in fermionic
liquids with impurities one arrives at the conclusion that the theory of
Anderson localization breaks down due to the presence of dissipation
(orthogonality catastrophe). In particular, no metal-insulator transition
is possible for heavy particles in $D=3$ since in any dimensionality
all states are localized.

Perhaps, a more important implication is for the interaction between
damped particles. The well-known effect of attractive short-range
interactions is superconductivity. In conventional superconductors
superconductivity is known to arise from Cooper pair formation by
electrons near the Fermi surface. For such electrons the density of states
can be taken constant. Therefore, the problem of Cooper pairing becomes
analogous (after a transformation to the center-of-mass and relative
coordinates) to single-particle localization on attractive short-range
potentials in $D=2$ where a single particle has constant density of
states. In the latter problem, the binding energy is exponentially
small:~\cite{Landau_Lifshits} $E\sim W\exp(-1/\alpha\nu)$, where $W$ is
the band-width, $\alpha$ is the strength of the potential and $\nu=m/2\pi$
is the single-particle density of states in $D=2$. This explains why the
critical temperature $T_c$ is so low in conventional superconductors.
While no single accepted theory of superconductivity in the
high-$T_c$ cuprate superconductors exists yet, most attempts to
explain high $T_c$ begin by considering strong interactions:
$\alpha\nu\sim 1$, while still following the BCS ideas (extended to the
strong-coupling case by Eliashberg). Here, I suggest that in systems where
the Landau Fermi liquid theory breaks down and the dynamics of fermions is
dissipative (governed by the orthogonality catastrophe) the nature of
Cooper pairing changes {\it qualitatively}. Most importantly, the binding
energy becomes much bigger and is essentially linear in $\alpha$ for weak
attractive interactions (see Eq.~(\ref{E_diss})). This means that any
weak second-order (fluctuation-induced) process is, in principle, capable
of producing sufficiently high $T_c$.

It should be emphasized, however,
that this model of superconductivity is a phenomenological one since the
arguments given in the beginning of the letter can not be used to deduce
the breakdown of the Landau theory. However, if the breakdown does occur,
this model is believed to capture the qualitative features of the physical
picture.

I would like to thank F. D. M. Haldane for sharing his insight
into the problem. I have also benefited from discussions with
L. P. Gor'kov and O. Motrunich. This work was supported by NSF MRSEC
DMR-9809483.


\end{document}